\documentclass[pre, preprint,amsmath,amssymbl, showkeys]{revtex4}

\usepackage{graphicx}
\usepackage{bm}
\usepackage{amssymb}

\begin{document}
%
\title{Self-guided wakefield experiments driven by petawatt class ultra-short laser pulses}

\author{S.~P.~D.~Mangles,
A.~G.~R.~Thomas,
C.~Bellei,
A.~E.~Dangor,
C.~Kamperidis,
S.~Kneip, 
S.~R.~Nagel, 
L.~Willingale and 
Z.~Najmudin}
\affiliation{Blackett Laboratory, Imperial College London, London SW7 2AZ, United Kingdom}

\begin{abstract}
We investigate the extension of self-injecting laser wakefield experiments to the regime that will be accessible with the next generation of petawatt class ultra-short pulse laser systems.
Using linear scalings, current experimental trends and numerical simulations we determine the optimal laser and target parameters, i.e. focusing geometry, plasma density and target length, that are required to increase the electron beam energy (to $> 1$~GeV) without the use of external guiding structures.
\end{abstract}

\maketitle

\section{Introduction}

\begin{figure}
\begin{center}
\includegraphics[width=4in]{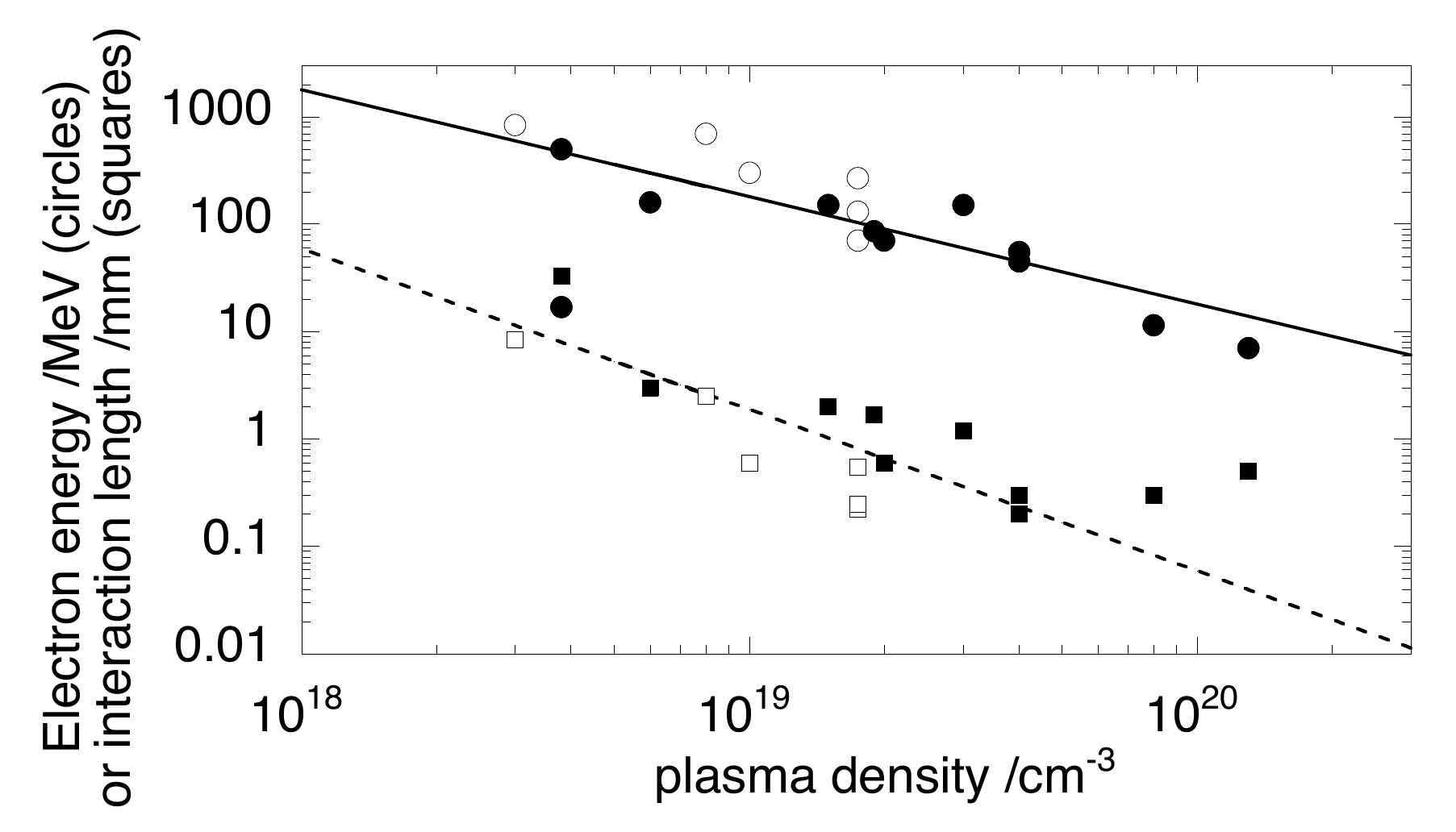}
\caption{Collation of reported data from various experiments and 3D particle-in-cell simulations.  
Closed circles: electron beam energy from experiments.  Open circles: electron beam energy from simulations. Closed squares: interaction length from experiments.  Open squares: interaction length from simulations. 
The solid line corresponds to equation \ref{Wmax} and the dashed line corresponds to equation  \ref{Ldp}. 
Data taken from \cite{Mangles:Nature, Geddes:Nature, Faure:Nature, Tsung:PRL2004, Reed:APL2006, Pukhov:APB2002, Miura:APL2005, Mangles:PRL2006, Leemans:NatPhys2006, Kotaki:LaserPhys2006, Hsieh:PRL2006, Hosokai:PRE2006, Hidding:PRL2006,  Hafz:NIMA2005, Geissler:NJP2006}}
\label{figure1}
\end{center}
\end{figure}

The field of laser driven electron acceleration has attracted significant international attention since the first observations of quasi-monoenergetic electron beams 
\cite{Mangles:Nature, Geddes:Nature, Faure:Nature}.
Results on the production of mono-energetic beams in self-injected laser wakefields from both experiments (solid symbols) and particle-in-cell simulations 
\cite{Tsung:PRL2004, Reed:APL2006, Pukhov:APB2002, Miura:APL2005, Mangles:PRL2006, Leemans:NatPhys2006, Kotaki:LaserPhys2006, Hsieh:PRL2006, Hosokai:PRE2006, Hidding:PRL2006,  Hafz:NIMA2005, Geissler:NJP2006} have been collated in figure \ref{figure1}. 

The maximum electron energy and acceleration length required in a laser wakefield accelerator can be derived from simple considerations.  The maximum length over which a wakefield can accelerate particles is typically governed by the dephasing length, that is the length over which a relativistic $(v \rightarrow c)$ electron will overtake the plasma wave, which travels at the group velocity of the laser pulse  $(v_g \simeq c(1-n_e/n_c)^{1/2} )$, by half the wavelength of a relativistic  plasma wave  $(\lambda_p = 2\pi c /\omega_p)$.  
 $n_e$ is the plasma density, $n_c = m\epsilon_0 \omega_0^2/e^2$ is the critical plasma density for electromagnetic wave propagation, where $\omega_0$ is the laser frequency ($n_c = 1.75 \times 10^{21}$~cm$^{-3}$ for 800~nm radiation).  The dephasing length for $n_e \ll n_c$ is given by the formula
\begin{equation}
\label{Ldp}
L_{dp} = \lambda_p \left( \frac{n_c}{n_e}\right)
\end{equation}

The maximum electron energy gain that a wakefield accelerator can produce can be estimated by integrating the electric field over a dephasing length (assuming a sinusoidal and non-evolving electric field). The electric field amplitude required to trap electrons that are initially at rest is given by $ E_0 =mc\omega_p / e$. 
Note that this is not the field at which 1D wave-breaking occurs, $E_{\rm{WB}} =  (2(\gamma_\phi -1))^{1/2}E_0$\cite{Akhiezer+Polovin}, where $\gamma_\phi$ is the Lorentz factor associated with the phase velocity of the plasma wave $(v_\phi \approx v_g$).  $E_{\rm{WB}}$ is the electric field at which the plasma wave can trap and accelerate an electron initially moving at $-v_\phi$  to one moving at $+v_\phi$. 
The field required to trap an electron at rest is more relevant to the transverse injection that is important for narrow energy spread beams.
This is because the electrons that are travel across the back of the first wave period, or `bubble', have low longitudinal momentum and can therefore be trapped at lower field strengths, typically on the order of $E_0$, than the 1D limit.  

This leads to an expression for the maximum energy an electron can gain from a plasma wave of amplitude $E_0$:
\begin{equation}
\label{Wmax}
W_{max} = 2 mc^2 \frac{n_c}{n_e}
\end{equation}
Equations \ref{Ldp} and \ref{Wmax} are also plotted on figure 1. 
It should also be noted that this figure is on a log-log scale, so that any scatter from the theoretical curve appears diminished.  
Nevertheless, the observed electron energies and interaction lengths appear to follow these simple scaling laws over a range of plasma densities.  
It should be noted that not all the data points are for similar laser systems.  
Data has been included from experiments with laser power as low as 2~TW and as high as 40~TW.  
Published 3D Simulation data has been included for laser powers over the range 10 - 300~TW.  
Both experimental and simulation data presented includes self-guiding and external guiding channel results (from less than 1 mm to $>$ 10 mm).

This collection of data clearly shows the general trend that to increase the electron beam energy experiments must move to lower plasma density and longer interaction lengths.  Stable 0.5 GeV acceleration has been achieved at LBNL following this approach, with some shots reaching the GeV level \cite{Leemans:NatPhys2006}.  
It should be noted that more detailed scalings were proposed in \cite{Gordienko:POP2005}  and \cite{Lu:PRST-AB2007} that include the dependence on the laser intensity but also assume the pulse dimensions are matched to the bubble or plasma wave dimensions which is not always the case in the experimental data presented.
These experiments typically rely on significant pulse evolution before electron injection can occur.

It is not sufficient simply to reduce the plasma density to increase the electron beam energy; the laser parameters must also be modified so as to reach this mono-energetic regime.  
The self-similar behaviour of these accelerators with density requires that the pulse dimensions scale with $\sqrt n_e$ while maintaining a minimum intensity; this requires the laser energy to increase with decreasing density.  
Tsung et al \cite{Tsung:PRL2004} and Lu et al \cite{Lu:POP2006, Lu:PRST-AB2007} have shown from simulations and analytical theory, that for injection at the back of the first wave period, a minimum intensity threshold of approximately $a_0 > 3$ is required.  

\begin{figure}
\begin{center}
\includegraphics[width=4in]{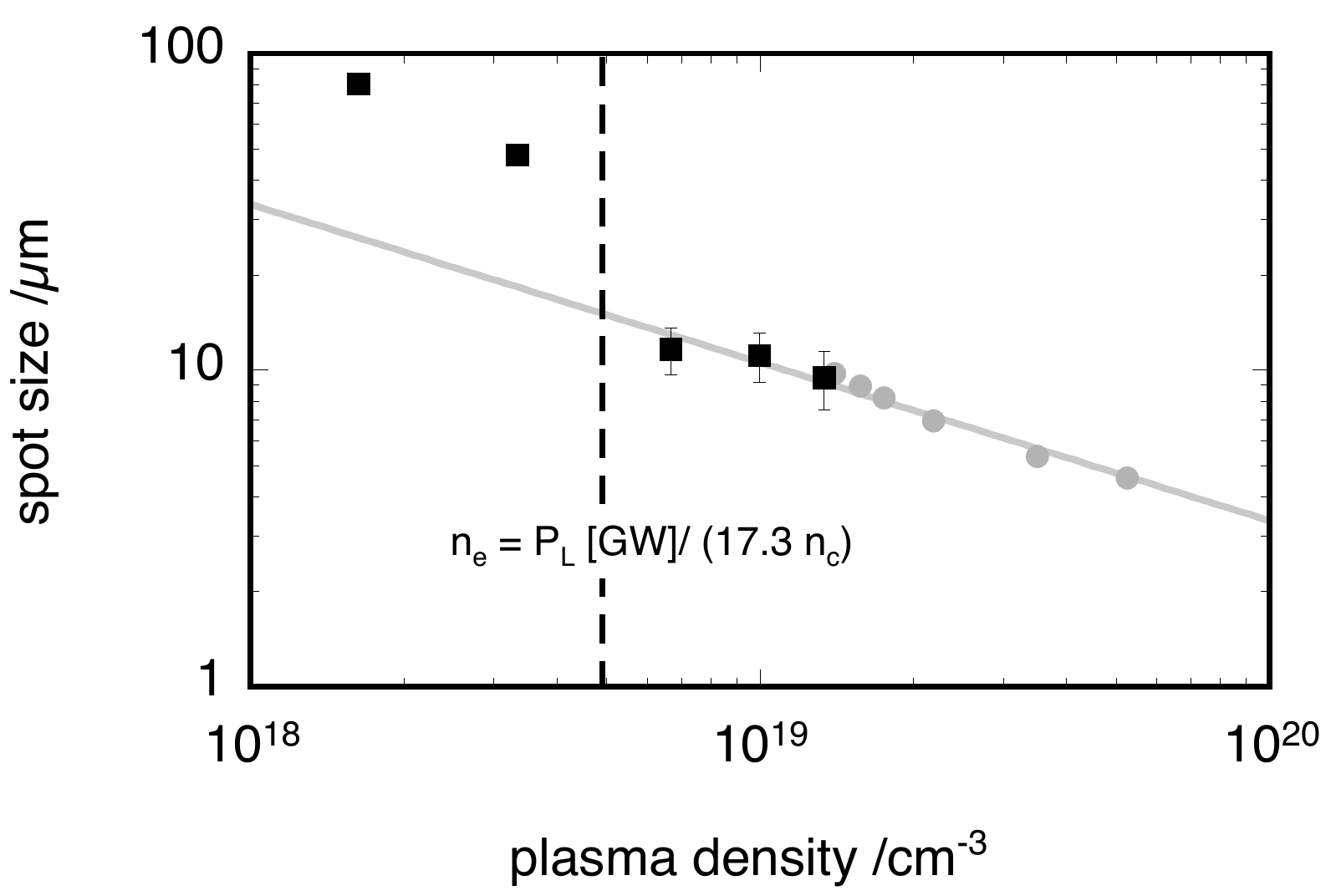}
\caption{Results from a recent Astra 15 TW experiment (black squares) and 2D PIC simulations (grey circles) showing the dependence on the laser spot size (FWHM intensity $\approx 1/e^2$ radius) with plasma density (experimental data taken from \cite{Thomas:PRL2007b}).  
The solid line indicates $\lambda_p  = 2 \pi c / \omega_p$.  The dashed line shows the density below which the laser in the experiment was below the critical power for self-focusing $P_c/ \rm{[GW]} = 17.3 (n_c/n_e)$. }
\label{Alecsdata}
\end{center}
\end{figure}

While pulse evolution (i.e. self-focusing and pulse compression) has hitherto played a crucial role in reaching self-injection with 10-100~TW lasers, it has also been attributed to some of the remaining shot-to-shot variability of the electron beam parameters.   
Experimental studies have shown that the stability of the electron beam is increased when the pulse and plasma parameters are chosen such that the beam waist $\rm{w}_0$ is matched to $\lambda_p$, minimising self-focusing effects  \cite{Mangles:POP2007} . 
It has also been experimentally verified that, in this regime, self-focusing tends to produce exit mode profiles with a beam waist approximately equal to the plasma wavelength \cite{Thomas:PRL2007b} for laser powers above the critical power for self-focusing. 
2D simulations also show that, for a range of plasma densities and focal geometries, self-focusing reduces the beam size until it is approximately $\rm{w}_0 \approx \lambda_p$ after which stable propagation occurs. 
Fig. \ref{Alecsdata} shows experimental and 2D Osiris simulation results that have been found to support this statement with 15~TW laser systems. 

A number of laser systems are now being constructed with laser powers 100 TW - 1PW and with laser pulse durations $ < 100$~fs.
One of the main goals of such lasers is their use as drivers for next generation particle acceleration experiments.
For example, a single beam of the Astra Gemini laser, which is under construction at the Rutherford Appleton Laboratory, is expected to produce a laser pulse of duration $\tau_L \approx 30$~fs with pulse energy of $\mathcal{E} \approx 15$~J on target \cite{Gemini_ref}.

To calculate the minimum density at which a laser such as Astra Gemini will be able to self-trap electrons we use a simple model \cite{Mangles:POP2007}.
We assume that the beam self-focuses to a matched spot size, $\rm{w}_0 \approx \lambda_p$ 
and require that the minimum intensity of the evolved pulse required to self-trap electrons  corresponds to a normalised vector potential of $a_t \approx 3.2$.
This value of $a_t$ was taken from experimental observations of the trapping threshold on 20~TW experiments \cite{Mangles:POP2007} and in agreement with the simulations reported in \cite{Tsung:PRL2004}.
The relationship between the laser pulse energy $\mathcal{E}_L$, pulse duration (FWHM) $\tau_L$, and plasma density $n_e$ can be found using by comparing the power of the laser pulse:
\begin{equation}
\label{ }
P_L = \frac{\pi \epsilon_0 m^2c^3\omega_0^2}{ 4 e^2} a_0^2 \rm{w}_0^2
\end{equation}
and the critical power for self-focusing: 
\begin{equation}
\label{ }
P_c = \frac{ 8 \pi \epsilon_0 m^2c^5}{e^2}\left(\frac{n_c}{n_e}\right)
\end{equation}
which produces
\begin{equation}
\label{ }
\frac{P_L}{P_c} = \frac{\pi^2}{8} \frac{a_0^2 \rm{w}_0^2}{\lambda_p^2}
\end{equation}
Inserting the threshold vector potential required for trapping, $a_t$ and using the fact that the peak laser power $P_L \approx 0.9 \mathcal{E}_L/\tau_L$ for a gaussian temporal profile, and that the spot size evolves towards $\rm{w}_0 \approx \lambda_p$ yields the result 
\begin{equation}
\label{energy-thres}
\frac{\mathcal{E}_{t}}{[\rm{J}]} = 23.3 \times 10^9 \left(\frac{n_c}{n_e}\right) a_t^2 \frac{\tau_L}{[\rm{s}]}
\end{equation}

\begin{figure}
\begin{center}
\includegraphics[width=4in]{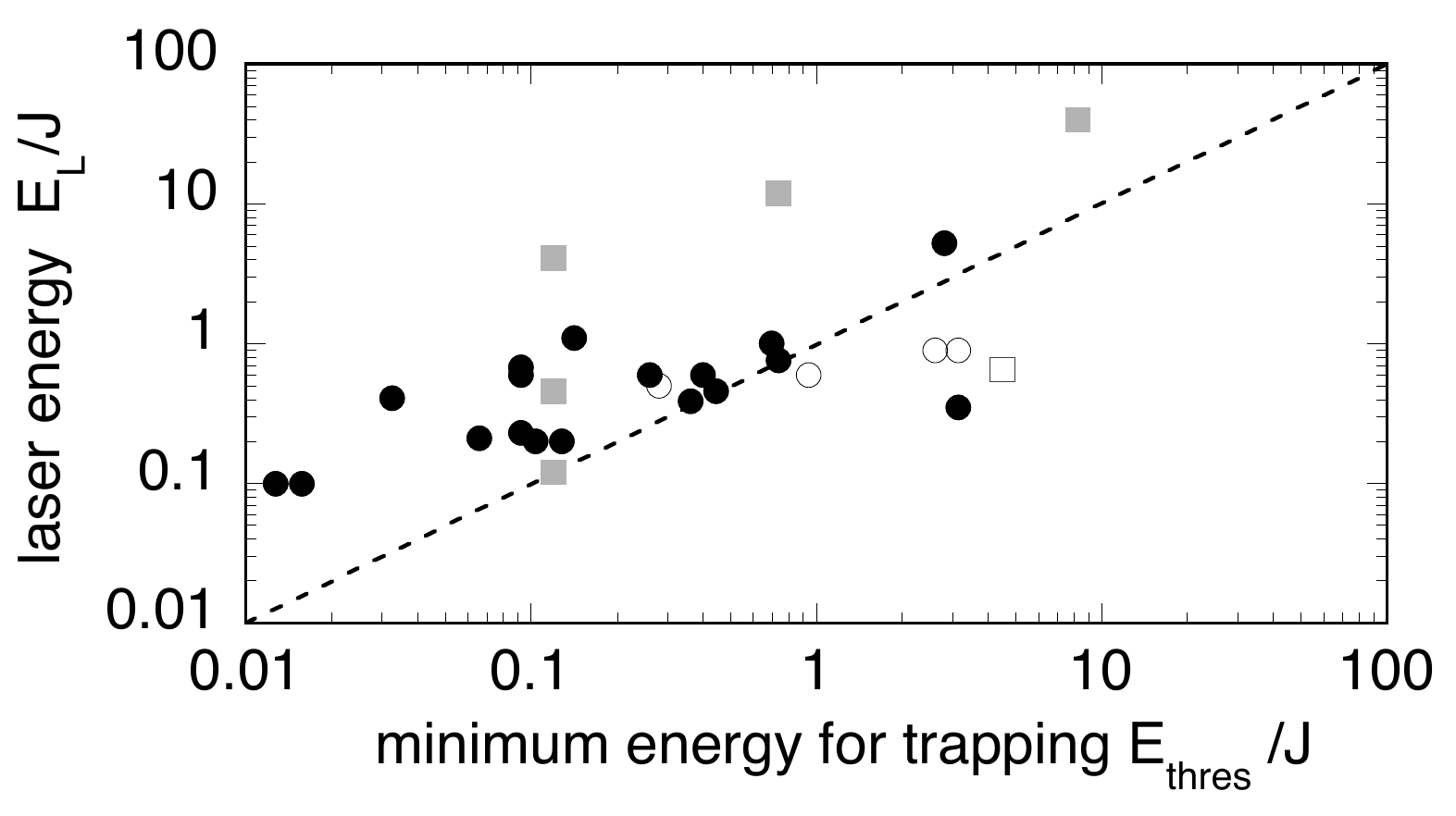}
\caption{Calculated minimum laser pulse energy required for self-trapping from equation \ref{energy-thres} for reported experimental and simulation data against the actual laser energy.  Solid circles: self-guided experiments. Solid squares: self-guided simulations. Open circles: guided experiments. Open squares: guided simulations.  The dashed line represents $\mathcal{E}_L = \mathcal{E}_{t}$ calculated for $a_t = 3.2$.  Data taken from \cite{Mangles:Nature, Geddes:Nature, Faure:Nature, Mangles:PRL2006, Mangles:POP2007,Mori:PhysicsLettA2006, Masuda:POP2007, Taki:AAC2006, Hidding:PRL2006, Miura:APL2005, Hsieh:PRL2006, RowlandsRees:Private, Leemans:NatPhys2006, Hosokai:PRE2006, Reed:APL2006,Ohkubo:PRST-AB2007, Masuda:JournalPhysiqueIV, Kotaki:LaserPhys2006, Tsung:PRL2004, Pukhov:APB2002, Hafz:NIMA2005, Geissler:NJP2006}}
\label{energy-thres-data}
\end{center}
\end{figure}

To verify this simple model we calculated the minimum laser energy required to achieve trapping for a range of reported experiments and simulations 
\cite{Mangles:Nature, Geddes:Nature, Faure:Nature, Mangles:PRL2006, Mangles:POP2007,Mori:PhysicsLettA2006, Masuda:POP2007, Taki:AAC2006, Hidding:PRL2006, Miura:APL2005, Hsieh:PRL2006, RowlandsRees:Private, Leemans:NatPhys2006, Hosokai:PRE2006, Reed:APL2006,Ohkubo:PRST-AB2007, Masuda:JournalPhysiqueIV, Kotaki:LaserPhys2006, Tsung:PRL2004, Pukhov:APB2002, Hafz:NIMA2005, Geissler:NJP2006}. 
The calculated energy threshold expression was modified to include the effect of pulse compression by replacing $\tau_L$ in equation \ref{energy-thres}  by $\pi /\omega_p$ (i.e. $ c\tau_L = \lambda_p / 2)$.  In one case \cite{Faure:Nature} we used the measured pulse duration after compression \cite{Faure:PRL2005}.
In figure \ref{energy-thres-data} we plot the calculated minimum energy required for self-trapping against the actual energy in the laser pulse.  
Note that only one self-guided experimental result \cite{Kotaki:LaserPhys2006} and no self-guided simulation results have a laser energy lower than the predicted threshold.
This implies that the energy threshold model is reasonably accurate for self-guided experiments.  
Apart from one experimental self-guided data point, all the other points that have $\mathcal{E}_L < \mathcal{E}_t$ are experiments or simulations where pre-formed guiding structures were used.  
This could indicate that the trapping mechanism in some guided experiments is not simply self-trapping or it could indicate that the presence of a radial density profile lowers the required intensity in the self-focused and compressed pulse as discussed in \cite{Lu:PRST-AB2007}.

\begin{figure}
\begin{center}
\includegraphics[width=4in]{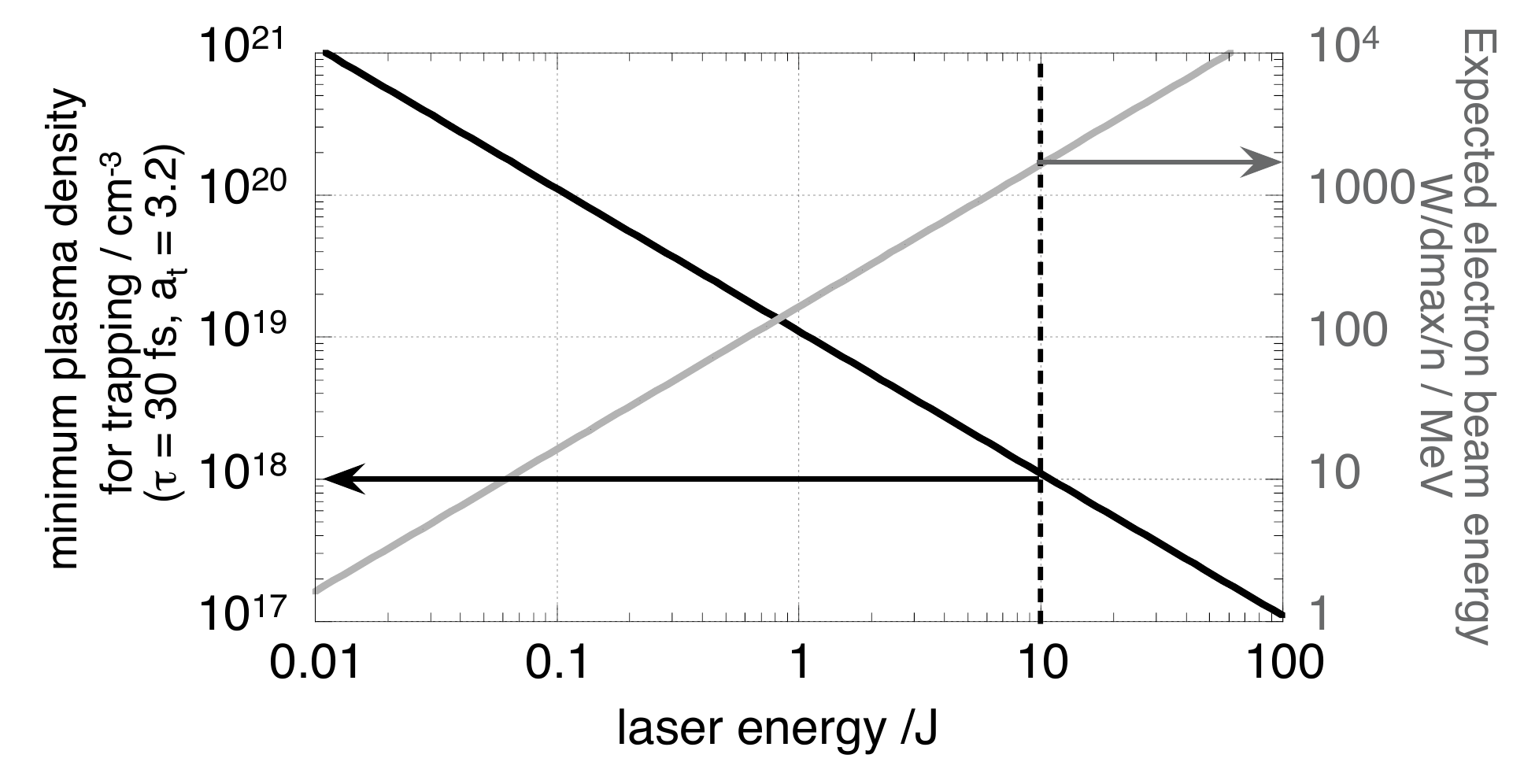}
\caption{Minimum density at which trapping will occur as a function of laser energy (left axis) and the expected eletron beam energy (right axis).  The threshold was calculated using equation \ref{energy-thres} with $\tau = 30$~fs and $a_t = 3.2$.  The dashed line indicates the position of a $\mathcal{E}_L = 10$~J laser (e.g. Astra Gemini).}
\label{laser-energy-trapping-density}
\end{center}
\end{figure}

The minimum density at which we expect trapping to occur (from equation  \ref{energy-thres} ) is shown as a function of laser energy for $\tau_L = 30$~fs in figure \ref{laser-energy-trapping-density}. 
For $\mathcal{E}_L =  10$~J we therefore expect the minimum density at which self-injection will be possible will be $n_e \approx 1 \times 10^{18}$~cm$^{-3}$ and that the corresponding electron energy will be $W_{\rm{max}} \approx 1.5 - 2$~GeV. 
The dephasing length at this density from equation \ref{Ldp} is close to 6 cm, which would require significant self-guiding.

\section{Numerical Modelling of Gemini electron acceleration experiments}

A series of simulations investigating potential electron acceleration experiments have been carried out for a $\tau_L = 30$~fs, $E_L = 10$~J $\lambda_L = 800$~nm laser system.
The simulations were performed using the particle in cell code Osiris \cite{Fonseca} in 2D3V slab geometry.  In these runs the laser propagates in the $x$ direction, the slab lies in the $x-y$ plane and the fields and particle momenta have components in $x$, $y$ and $z$. The laser was polarized in the $x-y$ plane. 
We use stationary ions. 
The simulations were performed on the 48 node `Caesar' cluster and on the CX1 supercomputer at Imperial College.
The simulation resolution was carefully chosen to minimise numerical dispersion errors while maintaining an acceptable run size. 
Typical run parameters were $\Delta x = 0.2~c/\omega_0$ ; $\Delta y = 0.8~c/\omega_0$ ; $\Delta t = 0.199~1/\omega_0$ in a simulation box size of  up to $(1600  \times 1600)~c/\omega_0 \approx 200 \times 200$~$\mu$m.  
The simulation box moves in the direction of the laser propagation at the speed of light.  
Runs were performed for propagation distances as large as 1~cm.  
The physical parameters of the runs presented in this report are shown in table \ref{table}.

The first 4 simulations were performed  at the minimum plasma density for self-trapping, chosen using equation \ref{energy-thres}, of $n_e = 1 \times 10^{18}$~cm$^{-3}$ but varying the spot size.  
In experiments the focusing geometry is not a simple parameter to change, and in PW class systems the necessary increase in beam size ($\sim 10$~cm diameter) to prevent damage to optics increases the size and cost of vacuum systems.  
Long focal lengths further exacerbate this situation and it is therefore advantageous to study the minimum focal spot size at which self-guiding will occur at the expected minimum density for self-trapping.

\begin{table}
  \centering 
  \begin{tabular}{ccccc}
\hline
Simulation  & density & Beam waist & Rayleigh  Range & vector  \\
 number &  /  cm$^{-3}$ &  $\rm{w}_0$  / $\mu$m  & $Z_R$ / mm  &   potential  $a_0$ \\ 
  1 & $1.05 \times 10^{18}$ & 5 & 0.1 & 18.8\\
  2 &  $1.05 \times 10^{18}$ & 10 & 0.4 & 9.6 \\
  3 & $1.05 \times 10^{18}$& 20 & 1.6 & 4.8 \\
  4 & $1.05 \times 10^{18}$& 34 & 4.5 & 2.8 \\
  5 & $2.10 \times 10^{18}$ & 20 & 1.6  & 4.8 \\
\hline
\end{tabular}
  \caption{Physical parameters used for the simulations presented. $\tau_l = 30$~fs (FWHM) and $E_L = 10$~J for all the simulations.}\label{table}
\end{table}

\begin{figure}
\begin{center}
\includegraphics[width=5in]{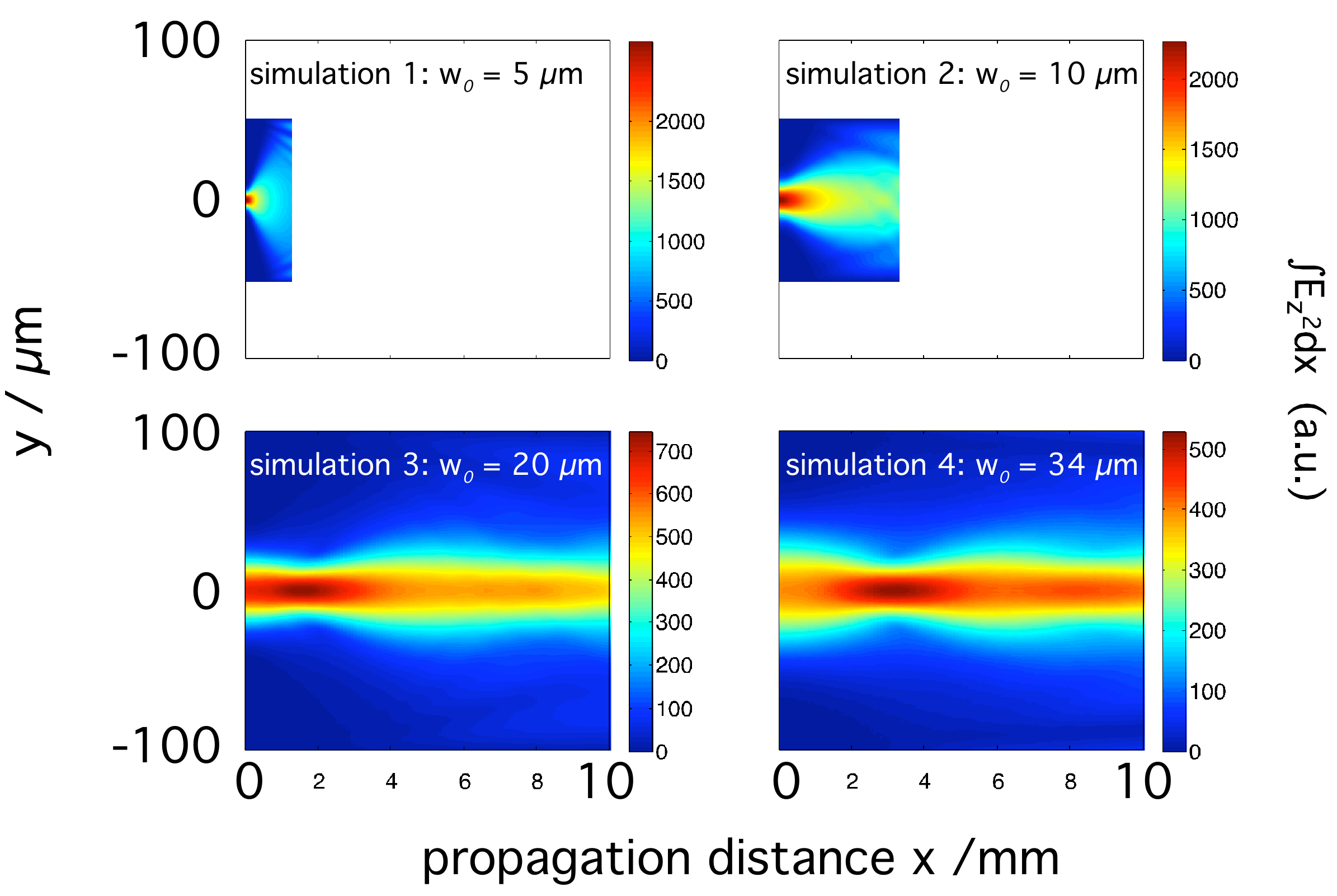}
\caption{Evolution of the transverse laser pulse envelope at a plasma density of $n_e = 1 \times 10^{18}$~cm$^{-3}$ for various beam waists. Simulations 1-4 are shown.  The transverse size of simulations 3 and 4 was increased  to 200~$\mu$m compared to 100~$\mu$m for simulations 1 and 2.}
\label{waist_evolution}
\end{center}
\end{figure}

Figure \ref{waist_evolution} shows the evolution of the laser electric field envelope in simulations 1 - 4.  
Each vertical slice in the image corresponds to the transverse profile of the laser envelope after integration along the propagation direction, $x$.   
For small spot sizes, $\rm{w}_0 = 5$ and $10~\mu$m, which correspond to $\rm{w}_0  < \lambda_p$, the pulse clearly diverges and is not significantly self-guided.
Indeed these simulations were halted after 2 and 4~mm propagation due to interaction of the diffracting pulse with the box boundary.  
As the spot size becomes close to the plasma wavelength guiding is observed. 
For $\rm{w}_0 = 20$ and $34$~$\mu$m there is still significant laser intensity ($a_0 \approx 2- 3$) after a propagation distance of 1 cm.  
The $\rm{w}_0 = 34$~$\mu$m simulation has guided over approximately $2Z_R$ ($Z_R = \pi\rm{w}_0^2/\lambda_L$ is the Rayleigh range) before the simulation was stopped.  
However the $\rm{w}_0 = 20$~$\mu$m case has guided for over $6Z_R$.

\begin{figure}
\begin{center}
\includegraphics[width=5in]{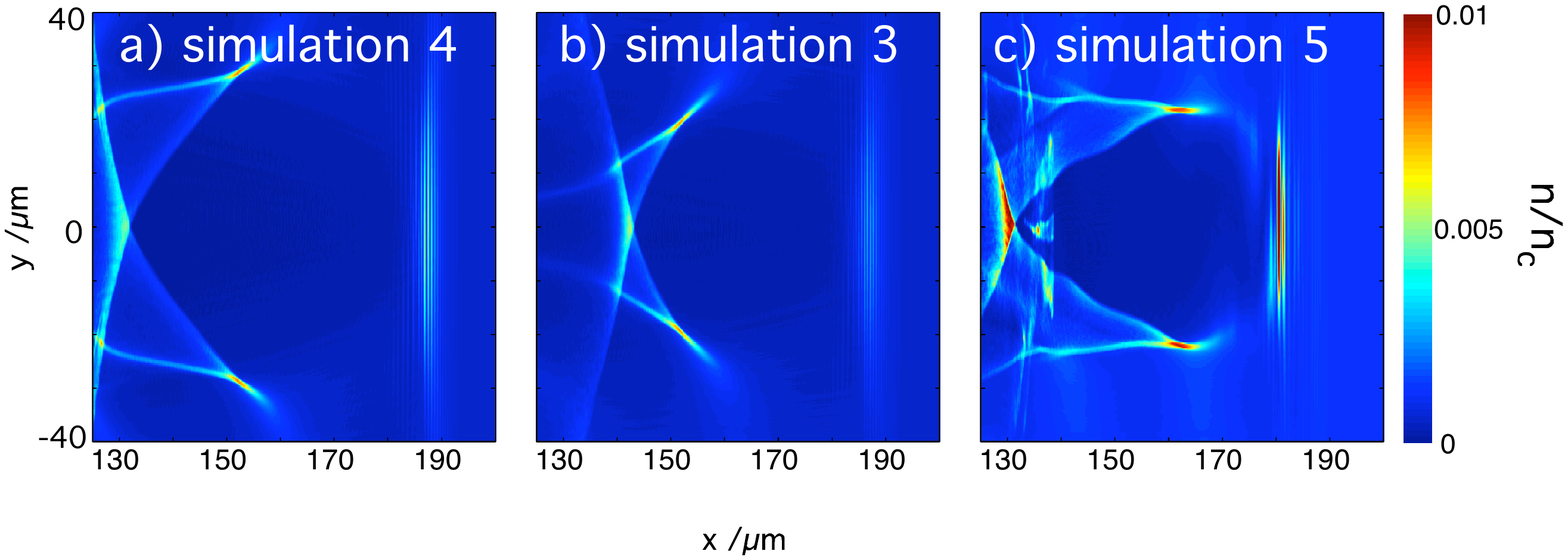}
\caption{Electron density profile after 9.6~mm propagation from three simulations investigating injection threshold.  The laser travels from left to right.}
\label{electron-density}
\end{center}
\end{figure}

Self-guiding alone is insufficient to produce an electron beam.  
The self-guided pulse must maintain a sufficiently high intensity to produce self-injection.  
In the following we concentrate on self-injection in the first wave period since  2D simulations can over estimate self-injection in the trailing periods compared with full 3D runs \cite{Tsung:Private}.  
In fig.~\ref{electron-density}  we show the electron density distribution of the first plasma wave period from three simulations (numbers 4,3 and 5 in table 1) after the laser has propagated for 9.6~mm.  In these plots the laser propagation direction is left to right.  

Simulation 4, at a density of $n_e = 1 \times 10^{18}$~cm$^{-3}$ and a spot size matched to the plasma wavelength $\lambda_p$ has not injected in the first period even at this late stage. In simulation 4, $a_0 = 2.8$, which is close to, but just below, the expected threshold for injection ($a_t \approx 3$).  
Pulse modification would be required before injection could occur, however as the pulse waist is already matched to $\lambda_p$ and the pulse length ($c\tau = 9$~$\mu$m) is less than $\lambda_p/2$ we might expect minimal self-focusing and compression.   

Simulation 3, also at a density of $n_e = 1 \times 10^{18}$~cm$^{-3}$ has also failed to produce self-injection in the first period.  
In this case $a_0 = 4.8$ which is above the expected threshold.  
This failure is explained by the fact that  in simulation 3 $\rm{w}_0 = 20$~$\mu$m,  which is less than $\lambda_p = 34$~$\mu$m so that some diffraction occurs initially, reducing the intensity before self-guiding occurs and so the plasma wave does not reach a sufficiently large amplitude.  

Simulation 5 shows the effect of moving to a higher plasma density ($n_e = 2 \times 10^{18}$ cm$^{-3}$), while maintaining the spot size of simulation 3. 
In this case the plasma wavelength is $\lambda_p  = 23$~$\mu$m, which is very close to the spot size $\rm{w}_0 = 20$~$\mu$m, and hence sufficient laser energy is trapped in the first plasma wave period and the intensity was high enough to cause self-injection after 4.2~mm propagation (i.e. some pulse evolution was required before injection occurred). 
The electron density plot shown in figure \ref{electron-density}  is after 9.6~mm of propagation, and therefore shows an electron bunch that has been accelerated for over 5~mm.
The fact that some pulse evolution was required prior to injection may indicate that the most stable beams from such a laser system may be achieved at a slightly higher plasma density, and therefore with a slightly lower electron beam energy.

\begin{figure}
\begin{center}
\includegraphics[width=2in]{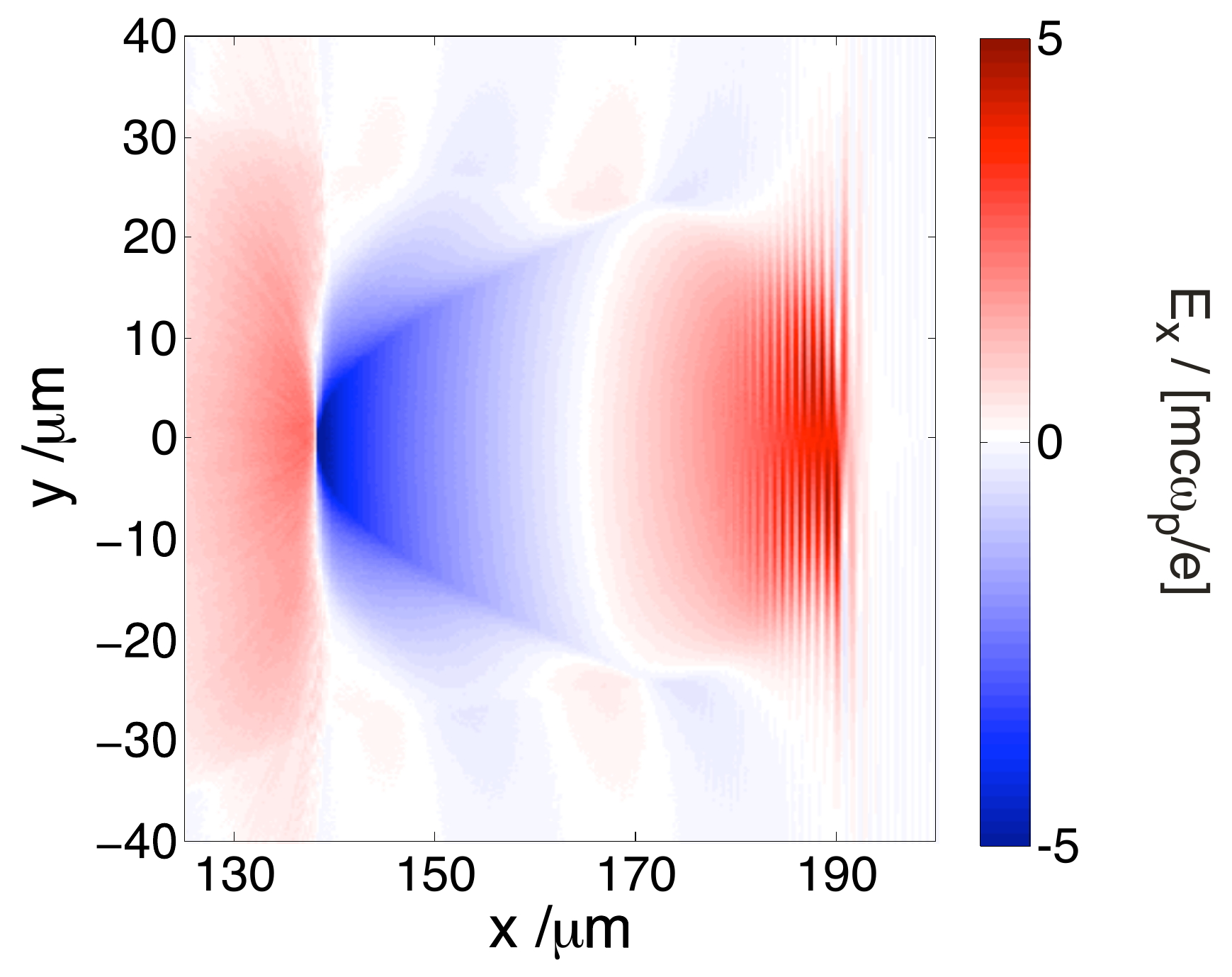}
\caption{Longitudinal electric field ($E_x$) of the first plasma wave period in simulation 5 after 4.2 mm propagation.  Blue (red) regions correspond to accelerating (decelerating) fields.}
\label{e-field}
\end{center}
\end{figure}

FIg.~\ref{e-field} shows the accelerating field at the point where injection has just occurred.  The peak accelerating field is close to $E_x = 5~mc\omega_p/e = 0.7$~GeV/mm. We therefore expect that, if this field were maintained over the entire acceleration length of 5~mm then electron energies close to 3 GeV would be obtained. This electric field at the back of the bubble is above the trapping threshold for stationary electrons, $E_0$, but is significantly less than the 1D cold wavebreaking field for this plasma density, $E_{\rm{WB}} \approx 40 E_0$, indicating that multi-dimensional effects are significant in determining the exact electron trajectories.  
That the field in the bubble is significantly larger than $E_0$ is somewhat at odds with the scaling presented in fig.~\ref{figure1}. Using that scaling we would predict peak electron energies of 0.85~GeV at the plasma density of simulation 5 ($n_e = 2 \times 10^{18}$~cm$^{-3}$).

\begin{figure}
\begin{center}
\includegraphics[width=4in]{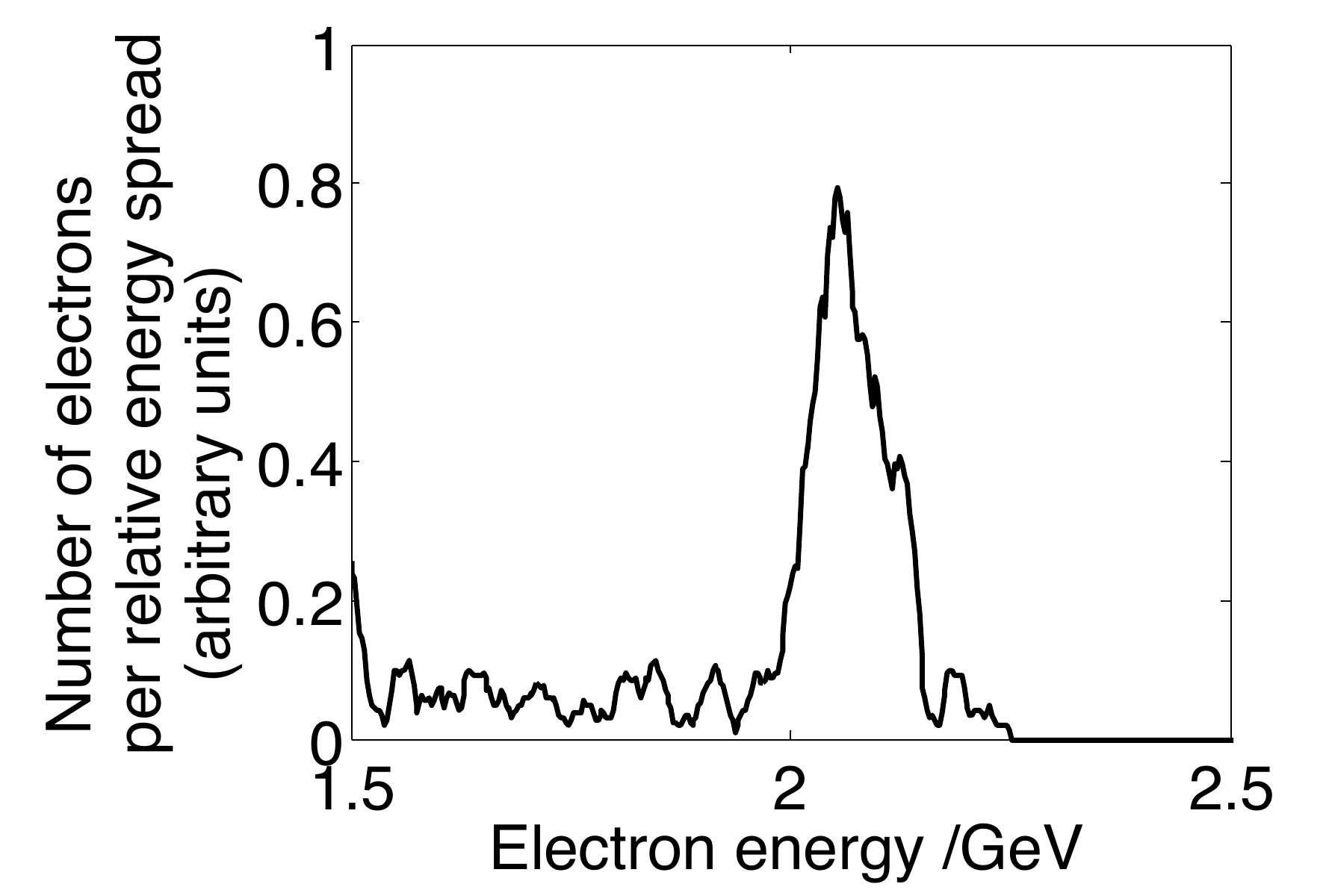}
\caption{Electron energy spectrum (number of electrons per relative energy spread) after 9.6~mm propagation in simulation 5.  The electron spectrum is calculated for electrons travelling within a cone angle of 25~mrad. The FWHM energy width of this beam is $\approx$ 2\%.}
\label{spectrum}
\end{center}
\end{figure}

An electron spectrum obtained from simulation 5, considering only electrons that would pass through an electron spectrometer with a 25 mrad acceptance cone (corresponding approximately to a typical electron spectrometer) is shown in fig.~\ref{spectrum}. This shows a quasi-mononenergetic spectrum at approximately 2 GeV (2 \%  relative energy spread) indicating that the size accelerating field does decrease slightly over time, possibly due to laser pump depletion or beam loading effects.  
We note that in this case the maximum energy is close to $W_{max} \approx 2 a_0 mc^2 (n_c/n_e)$, which is consistent with 3D non-linear scalings including the dependence on $a_0$ \cite{Lu:PRST-AB2007}. 

\section{Summary}

In a series of particle-in-cell simulations we have investigated the parameters of interest for the next generation of experiments on self-guided laser wakefield acceleration.  
Using simple scaling laws we identified the regions of interest and have then verified and refined these using 2D PIC simulations.  
We have considered the appropriate focusing geometry and plasma target requirements necessary to produce quasi-monoenergetic electron beams to the multi-GeV level.  
The simulations show that a focal spot size on the order of 20~$\mu$m (i.e. $f/20$ focusing) is ideal and that plasma targets of centimetre length capable of producing densities up to a few $10^{18}$~ cm$^{-3}$ are suitable.  
This indicates that multi-GeV beams from a self-guided wakefield accelerator should be experimentally realisable in the near future.


\section*{Acknowledgment}
This work was funded by EPSRC. S.P.D.M. thanks the Royal Society for support.
We gratefully acknowledge the Osiris consortium (UCLA/IST/USC) for the use of the Osiris particle-in-cell code.

\bibliographystyle{unsrt}
\bibliography{Mangles_Gemini_sims}

\end{document}